# National Oceanic and Atmospheric Administration Publishes Misleading Information on Gulf of Mexico "Dead Zone"

Michael W. Courtney and Joshua M. Courtney

BTG Research, P.O. Box 62541, Colorado Springs, CO, 80962
Michael_Courtney@alum.mit.edu

**Abstract**
Mississippi River nutrient loads and water stratification on the Louisiana-Texas shelf contribute to an annually recurring, short-lived hypoxic bottom layer in areas of the northern Gulf of Mexico comprising less than 2% of the total Gulf of Mexico bottom area. Many publications demonstrate increases in biomass and fisheries production attributed to nutrient loading from river plumes. Decreases in fisheries production when nutrient loads are decreased are also well documented. However, the National Oceanic and Atmospheric Administration (NOAA) persists in describing the area adjacent to the Mississippi River discharge as a "dead zone" and predicting dire consequences if nutrient loads are not reduced. In reality, these areas teem with aquatic life and provide 70-80% of the Gulf of Mexico fishery production. On June 18, 2013, NOAA published a misleading figure purporting to show the "dead zone" in an article predicting a possible record dead zone area for 2013 (http://www.noaanews.noaa.gov/stories2013/20130618_deadzone.html). This area is not a region of hypoxic bottom water at all nor is it related directly to 2013 predicted hypoxia. This figure appeared as early as 2004 in a National Aeronautics and Space Administration (NASA) article (http://www.nasa.gov/vision/earth/environment/dead_zone.html) as a satellite image where the red area represents turbidity and is much larger than the short-lived areas of hypoxic bottom water documented in actual NOAA measurements. Thus, it is misleading for NOAA to characterize the red area in that image as a "dead zone." The NOAA has also published other misleading and exaggerated descriptions of the consequences of nutrient loading.

**Introduction and Background**
The authors have been following the literature on nutrient loading and hypoxia in the Gulf of Mexico (Bianchi et al. 2010, Rabalais et al. 2007, Rabalais et al. 2002) for several years. Characterizations of these areas as "dead zones" are perplexing, and the Louisiana-Texas shelf teems with life. Louisiana harvests tremendous quantities of seafood and supports one of the best sport fisheries in the Gulf of Mexico. Our anecdotal observations and review of available data suggest that several sport species of fish (red drum, spotted sea trout, red snapper) are both plumper and more plentiful in Louisiana waters, and we have often considered hypothetical explanations for these observations. In 2011, the last time a record "dead zone" area was predicted, (Thean 2011, Rice 2011) we were actively conducting creel surveys in the Port Fourchon and Calcasieu areas of the Louisiana Gulf Coast to study the impact of the Deepwater Horizon oil spill. (Courtney et al. 2012) These areas showed some slight impacts from the oil spill near Port Fourchon and the effects of overfishing in the Calcasieu area, but these areas were far from being accurately characterized as "dead zones."

The area of the Louisiana-Texas shelf shown as a "dead zone" in the figure accompanying the NOAA report of 18 June 2013 (NOAA 2013) is well known for the recent resurgence and impressive growth in the population of red snapper (*Lutjanus campechanus*). This population recovery has been widely attributed to oil platforms and other artificial reefs (Shipp and Bartone 2009, Gallaway et al. 2009). While reviewing Melissa Monk's doctoral thesis (Monk 2012, Figures 1.14 – 1.22), we noticed the apparent decrease in faunal biomass (SEAMAP trawl surveys) with increasing distance from the Mississippi River discharge as well as the greater animal biomass in the area with higher nutrient loading and seasonal bottom water hypoxia. Monk's suggestion that Mississippi River outflow and nutrient loading should be considered to explain the temporal variability in both total biomass and how the biomass is distributed among species led us to formulate and assess the literature for evidence supporting the hypothesis that nutrient loading from the Mississippi River also has contributed significantly to the recovery of the red snapper population (Courtney et al. 2013) in the regions shown in red in the figure from the NOAA article (also Figure 1, present paper).

The highest annual harvests of red snapper in the Gulf of Mexico before 1970 were approximately 9 million pounds (mp), and in most years were about 5 mp (Shipp and Bartone 2009). These yields were not sustainable at the time and led to a collapse of the fishery in the eastern Gulf between 1980 and the early 1990s. However, current model projections of maximum sustainable yield of red snapper for the Gulf are between 11.3 and 25.4 mp annually, with production dominated by the northern and western Gulf, especially the Louisiana-Texas shelf (Shipp and Bartone 2009, Cowan et al. 2011). Understanding that this area supports such abundant marine life, the NOAA characterization of the area as a "dead zone" motivated a more careful look.

**Bottom Water Hypoxia and Annual Dead Zone Survey**

Since the region of bottom water hypoxia is annually recurring most summers, the NOAA conducts annual surveys to measure dissolved oxygen concentrations. LUMCOM and Texas A&M conduct additional surveys. With so much survey data available to them, and being familiar with their annual survey maps, the appearance of a figure depicting a much larger "dead zone" area (NOAA 2013) was a surprise. Investigating the source of this image revealed that it is not actually a "dead zone" at all but rather an image of satellite data showing turbidity which might be related to abundant phytoplankton from the NASA Aqua satellite's Moderate-Resolution Imaging Spectroradiometer (MODIS) instrument.

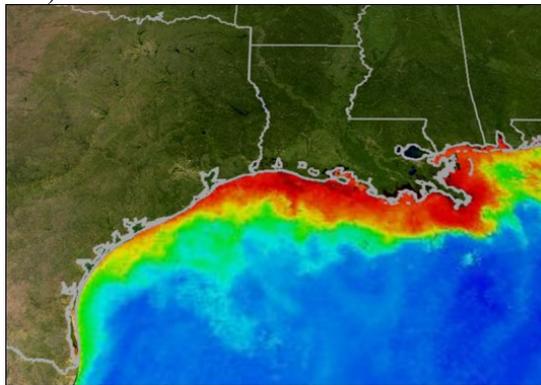

*Figure 1: (NOAA 2013) Figure published by NOAA in 2013 and credited to NOAA purporting to show "dead zone" in red. The region in red is not an area of bottom water hypoxia, but rather an area of high turbidity in satellite data published by NASA as early as 2004 (NASA 2004). Note: as a work of the US Government, this figure is not subject to copyright and no permission is needed to reproduce it.*

Images of actual bottom water oxygen measurements can be downloaded from http://www.ncddc.noaa.gov/hypoxia/products/. Figure 2 shows several for recent years 2008-2013. Note that the actual red areas in these figures (temporary bottom water hypoxia characterized by NOAA as "dead zones") are much smaller than the red area in Figure 1 that NOAA claims to be a "dead zone." Furthermore, there is no NOAA survey data supporting the inference from Figure 1 that inshore areas such as Lake Borgne, Chandeleur Sound, Black Bay, Timbalier Bay, Terrebonne Bay, or Vermillion Bay are regions with prevalent, annually recurring bottom water hypoxia. In contrast, these areas are popular locations for both recreational and commercial fishing, as are many of the offshore areas. The area closest to the mouth of the Mississippi River (often referred to as Venice) has some of the best fishing in the world.

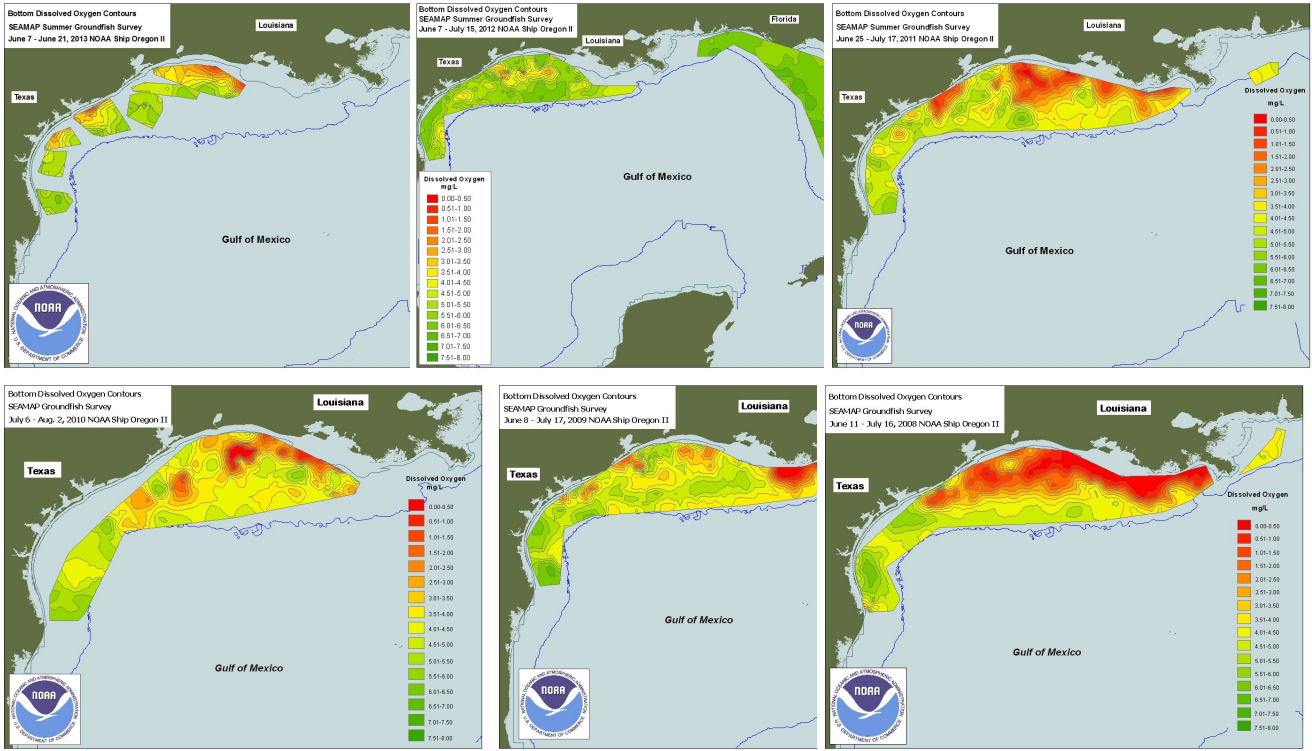

*Figure 2: Actual NOAA survey maps showing regions of short-lived bottom water hypoxia from 2008-2013. Note: as works of the US Government, these figures are not subject to copyright and no permission is needed to reproduce it.*

**Nutrient Enrichment Increases Fishery Production of Marine Systems**
Nixon and Buckley (2002) review evidence that secondary production in marine systems tends to be increased by nutrient enrichment. They cite the example of the collapse of the Egyptian shelf fishery after Aswan dam was closed in 1965. Restricting the deposition of inorganic nutrients by the annual Nile flood caused the fishery to collapse. Subsequently, use of inorganic fertilizers and point source discharge of nutrients increased markedly in the Nile delta, and the resulting increase in inorganic nitrogen coincided with a dramatic recovery of the fishery that began in the early to mid 1980s and has continued to the present. Caddy (2000) and Grimes (2001) also discuss the impact of the Aswan dam on the fisheries off the Nile delta and the later rise in production with enriched drainage water in the early to mid 1980s.

Nixon and Buckley (2002) also consider the Baltic Sea, which has received increasing amounts of nitrogen and phosphorus from agricultural runoff and other sources for many decades. Between an early assessment (1920-1923) and a later study (1976-1977), benthic biomass above the halocline was estimated to increase by 2-10 times. Decreases in benthic biomass below the halocline were attributed to hypoxia. Thurow (1997) conducted a historical reconstruction of both biomass and finfish yield in the Baltic Sea. Both biomass and finfish yield were low in the first half of the 20th century, then increased sharply after about 1950 with the increasing use of agricultural fertilizers. Thurow (1997) concluded that nutrient loading and resulting enhanced primary production were ultimately responsible for the 8-fold increase in fish biomass. While some argued for reduction of nutrient inputs to reduce areas of hypoxic bottom layers, Thurow (1997) warned that reduction of nitrogen and phosphorus inputs would "inevitably lead to a lower fish biomass."

Caddy (1993) has provided a model relating nutrient loading to fishery landings. Fishery landings respond positively to nutrient loadings up to some maximum, beyond which fishery landings decrease in response to excessive nutrient loads. Caddy hypothesized that higher levels of nutrient loadings (above a system dependent threshold) lead to seasonal or permanent bottom anoxia causing a decline in bottom dwelling organisms (benthos) and the fish that feed on them. Caddy (2000) also connected production increases in the northern Mediterranean with increased nutrient load, even though some areas (the northern Adriatic) have experienced occasional hypoxia. Oczkowski and Nixon (2007) reviewed available data and showed that fish landings in the Nile delta increase with nitrogen loading up to a threshold concentration of approximately 100 µM of dissolved inorganic nitrogen. However, above this threshold, there is a steep decline in fishery landings. Breitberg (2002) has also discussed how the high productivity of finfish in nutrient-enriched systems is often greater than losses due to hypoxia.

Churchill B. Grimes reviewed both the literature and the available data related to the Mississippi River's nutrient loading in the Gulf of Mexico (Grimes 2001). Grimes concluded that nutrient loading from the Mississippi River contributes significantly to fishery production in the areas adjacent to the Mississippi River discharge. All the available evidence suggests that the area of the Gulf of Mexico adjacent to the Mississippi River discharge is on the rising edge or near the peak of the Caddy curve in most years. In other words, if nutrient loading is substantially increased, fishery declines are a possibility only if the system is pushed beyond the peak in the system response. However, current nutrient loads and $20^{th}$ century increases of nutrient loading from the Mississippi River have significantly enhanced fishery production, and the most likely outcome of a significant reduction in nutrient loading would be a decline in production on the Louisiana-Texas shelf and in adjacent waters in the northern Gulf of Mexico off the Mississippi and Alabama coasts.

**Discussion**

When preparing scientific information for public discussion, there is the temptation to oversimplify to make science appealing and to enhance effective communication. Inconvenient facts may be glossed over in hopes of swaying public opinion and moving toward changes in public policy. In the discussion of temporary, annually recurring bottom water hypoxia in the Gulf of Mexico, we have noticed tendencies to emphasize the causal role of agricultural nitrogen and phosphorus and not to emphasize the role of point sources or the role of carbon released from mobile muds (Bianchi 2008). We have also noticed the tendency to emphasize worst case scenarios rather than the important roles of seasonal precipitation, stratification, wind forcing, and mixing. Past predictions (Rice 2011) that turned out to be cases of "chicken little" are forgotten, and the prognostications of this year's experts are taken as gospel, forgetting that real science is about the validation of predictive models through experiment and observation rather than making news and influencing public opinion through unvalidated predictions. The largest area of temporary bottom water hypoxia occurred in 2002 in spite of predictions regarding record breaking "dead zones" in 2008 and 2011. The history of unfulfilled predictions justifies a "wait and see" approach to a similar prediction in 2013.

Furthermore, it is misleading to continually compare the area of hypoxic zones to states such as New Jersey rather than with the area of the Gulf of Mexico or Louisiana-Texas shelf as a whole. On a percentage basis, hypoxic bottom areas in the Gulf of Mexico (< 2%) are much smaller than hypoxic areas of Lake Erie, the Black Sea, or Chesapeake Bay in years of significant hypoxic events.

Due care needs to be taken when selecting figures to make choices that accurately represent the physical quantities being attributed to them (dissolved oxygen rather than turbidity which might be phytoplankton). It is ironic that what might be evidence of life (phytoplankton) in Figure 1 is portrayed as a "dead zone." Primary production of phytoplankton is an essential part of marine food webs, and in most cases (including many cases of nutrient enrichment and eutrophication) significant increases in primary production lead to increased biomass and more productive fisheries. In spite of the attention grabbing moniker of "dead zone," the demonstrated impact on fisheries production of nutrient loading in the Gulf of Mexico has been generally positive, and negative impacts on fisheries landings are largely hypothetical.

Due care also needs to be taken not to mislead the readership when pairing a figure with an article. The NOAA (2013) article is about hypoxia predictions in 2013, so selecting a figure that appeared as early as 2004 is misleading, as is attributing the image to NOAA when it appears to have originated with NASA from the MODIS/Aqua satellite.

Finally, the NOAA (2013) claim demonstrates several large and unsupported leaps of logic. The satellite image actually documents turbidity which might be a phytoplankton bloom. However, everyone familiar with the area knows that two main contributors to turbidity for inshore and near shore Louisiana Gulf Coast waters are wave action stirring up bottom sediments (deposited over many years by the Mississippi River) and sediments carried downstream in the present year by the Mississippi River. The popular description of muddy water by local fishermen is "chocolate milk" as they hope for a break in the southerly winds and wave action so the sediments settle and reveal the "green" water they prefer. Other large and unsupported leaps of logic are that short-lived areas of hypoxic bottom water are more harmful than reducing nutrient levels to those believed necessary to achieve long term goals of bottom water hypoxic area.

Good science and science based news should provide a clear trail from scientific images back to the sources of data from which such figures and images were produced. Figure 3 shows chlorophyll a estimates from the MODIS/Aqua satellite as seasonal composites for winter 2003-2013. This figure was obtained from the NASA OceanColor web site (http://oceancolor.gsfc.nasa.gov/cgi/l3 ) at a 4 km resolution. Chlorophyll a is a common way to estimate primary production aquatic systems. Thus, it is an indicator of life and potential of a marine ecosystem rather than an indicator of a "dead zone." The similarity in areas between this (winter) view and the analogous summer image of chlorophyll a shows that the appearance of red in the summer is not a harmful algae bloom, but rather a relatively constant high level of photosynthesis (with normal seasonal variations) in a healthy and very productive fishery.

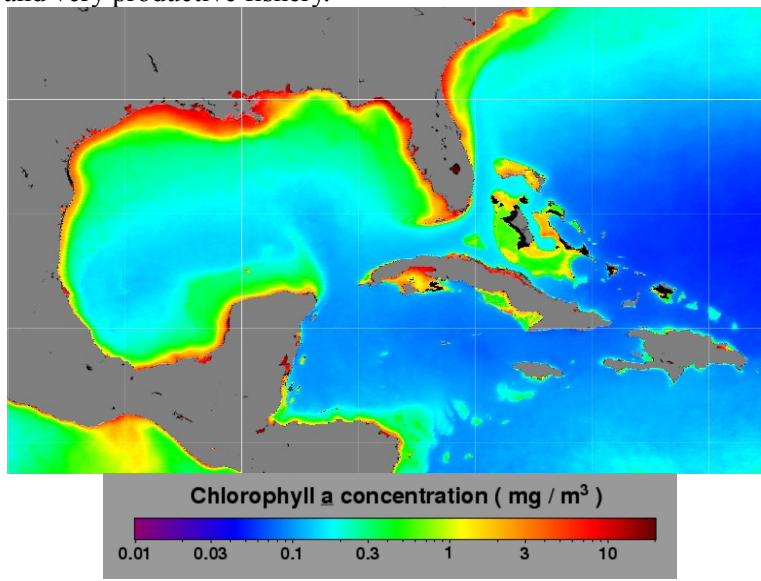

*Figure 3: Map of winter chlorophyll a concentrations from 2003-2013 as determined from MODIS instrument on NASA's Aqua satellite. Note: as a work of the US Government, this figure is not subject to copyright and no permission is needed to reproduce it.*

**Acknowledgments**
We thank Amy Courtney, PhD, for helpful discussion and comments that improved the manuscript.

**About the Lead Author**
Michael Courtney's first laboratory job was in Fisheries Science at the LSU Aquaculture facility in 1985-1986. He received the LSU University Medal in 1989 for graduating ranked first in his class with a BS in Physics before pursuing a PhD in Physics from MIT, completing that degree in 1995. He has been an active researcher in blast physics and ballistics since 2001, and served on the Mathematics faculty of the United States Air Force Academy from 2009 to 2013. He returned to research in Fisheries Science while directing the Air Force Academy's Quantitative Reasoning Center and has published numerous papers bringing quantitative analysis and scientific insights to bear on a variety of problems. He currently serves as a consulting scientist for BTG Research (www.btgresearch.org).